# Optical emission of graphene and electron-hole pair production induced by a strong THz field

*I.V. Oladyshkin\*, S.B. Bodrov, Yu.A. Sergeev, A.I.Korytin, M.D Tokman, and A.N. Stepanov*

Institute of Applied Physics of the Russian Academy of Sciences
46 Ul'yanov Street, 603950, Nizhny Novgorod, Russia

We report on the first experimental observation of graphene optical emission induced by the intense THz pulse. P-doped CVD graphene with the initial Fermi energy of about 200 meV was used, optical photons was detected in the wavelength range of 340-600 nm. Emission started when THz field amplitude exceeded 100 kV/cm. For THz fields from 200 to 300 kV/cm the temperature of optical radiation was constant, while the number of emitted photons increased several dozen times. This fact clearly indicates multiplication of electron-hole pairs induced by an external field itself and not due to electron heating. The experimental data are in a good agreement with the theory of Landau-Zener interband transitions. It is shown theoretically that Landau-Zener transitions are possible even in the case of heavily doped graphene because the strong THz field removes quasiparticles from the region of interband transitions during several femtoseconds, which cancels the Pauli blocking effect.

*\*oladyshkin@gmail.com*

The nonlinear optical properties of graphene are currently actively investigated in view of their prospective use in plasmonics, optoelectronics, and photonics [1]. The specific features of gapless dispersion of Dirac fermions, including the so-called Dirac cones, make graphene a unique material. In the neighborhood of the Dirac point, fermions have a massless dispersion law up to energies of order 1.5 eV and high Fermi velocity $\upsilon_F \approx 10^8$ cm/s, thus providing ultrahigh nonlinear susceptibility (both quadratic and cubic) of graphene in the infrared and terahertz ranges [2-10]. The interaction of graphene with terahertz radiation is presently arousing particular interest from the viewpoint of various applications [11]. Terahertz video pulses are also used for studying relaxation processes in graphene [12,13].

An interesting effect – the THz-pulse-initiated carrier multiplication (CM) was studied experimentally in [12]. This effect was attributed in [12] to impact ionization (II), i.e., to interband reverse Auger recombination. At the same time, according to the theoretical results obtained in [14] within the framework of the 2D Fermi-liquid model, the Auger processes have low efficiency in the region of the linear dispersion law[1]; consequently, in the simulation made in [12] the Auger resonances were broadened "artificially" within the framework of the numerical scheme[2]. The CM effect was also observed in the case of optical pumping [13, 16-22], when the contribution of the reverse Auger relaxation to the formation of electron population in the conducting band did not exceed 5-10% [17].

It is worth noting that the ballistic (directly initiated by the field) interband transitions were neglected in [12] in view of the small photon energy as compared to the typical kinetic energy of fermions in doped graphene. This approach, however, is justified only within the applicability of the perturbation method, when the energy of the quasi-particle interaction with an h.f. field is small compared to the energy of the resonance transition: $e\upsilon_F A/c \ll \hbar\omega$ (here A is the vector potential amplitude). The typical field intensity $E \approx$ 100 kV/cm for the frequency of 2 THz reported in [12] by no means corresponds to the given parameter region $e\upsilon_F A/c\hbar\omega \approx 30$.

In our work, the ballistic CM mechanism under the action of intense THz pulses is studied experimentally and theoretically. We address the mechanism of Schwinger electron-hole pair production earlier considered in connection with transport processes in graphene [23-25]. This effect may be also adequately described within the framework of the Landau-Zener theory [26-28]. In dc fields, this effect usually makes a relatively small contribution to the current-voltage characteristic of a graphene sample that is observed at weak doping only [26,27]. We have derived a conclusion that, in the field of intense THz pulses, the ballistic mechanism of electron-hole pair generation, contrariwise, plays a principal role in the formation of fermion distribution, even in the case of strong initial doping in CVD graphene, when the initial Fermi energy is of order hundreds of meV.

The architecture of the paper is the following. **Section 1** is devoted to the Landau-Zener mechanism of ballistic electron-hole pair generation. The experiment on excitation of spontaneous optical emission of graphene by high-power THz radiation[3] is described in **Section 2**. In **Section 3** the dependence of the intensity of optical emission on the field amplitude of the THz pulse obtained in experiment is explained within the Landau-Zener theory based on the hypothesis of the decisive role of ballistic interband transitions.

---

[1] The situation is different for graphene in magnetic field (see, e.g., [15]).
[2] The authors of [12] assumed that dislocations could be a possible physical mechanism of such broadening.

**1. Theoretical model of ballistic ionization**

The fermion energy spectrum in the neighborhood of the Dirac point has the form [1]:

$$W_{c,\upsilon} = \pm \hbar \upsilon_F \sqrt{k_x^2 + k_y^2}, \quad (1)$$

where $\hbar k_{x,y}$ are the components of quasi-momentum in the single-layer plane, the $c$ and $\upsilon$ subscripts define the conducting and valence bands, respectively. Let us consider the $W(k_x)$ dependence at a given value of $k_y$ (Fig.1).

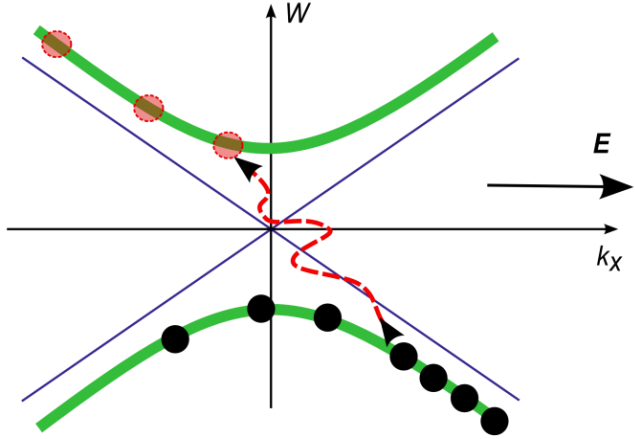

**Fig. 1.** The dependence of the Dirac fermion energy $W$ on quasi-momentum $k_x$ for $k_y$=const (Eq.(1)). The intersecting lines correspond to $k_y=0$. The dotted line depicts schematically the interband transition under the action of THz field.

For $k_y = 0$, the energy diagram is formed by two intersecting straight lines $W_\pm = \pm \hbar \upsilon_F k_x$, with the energy sign changing at the point $k_x = 0$ of each line. For $k_y \neq 0$, an energy gap having width $\Delta W_{min} = 2\hbar \upsilon_F |k_y|$ is formed between the conduction and valence bands.

Let the electric field $\mathbf{E} = \mathbf{x}_0 E(t)$ be imposed on the system. Changes of the quasi-momentum will be described by the classical equations of motion

$$\hbar \dot{k}_x = -eE, \quad k_y = const. \quad (2)$$

If the quantity $k_y$ is regarded to be a perturbation defining splitting of the energy branches (Fig.1) and $k_x(t)$ is a parameter specifying the difference between the energies of states at each moment of time, then we will obtain a standard Landau-Zener problem [31]. According to the general theory of Landau-Zener transitions, the probability of the interband transition is defined as $P_{LZ} = \exp(-2\pi\Gamma)$, $\Gamma = \hbar^{-1}(\Delta W_{min}/2)^2 |\partial(W_+ - W_-)/\partial t|^{-1}$, where the derivative $\partial(W_+ - W_-)/\partial t$ shall be calculated for unperturbed (intersecting) branches of the spectrum. In our case, $\partial(W_+ - W_-)/\partial t = 2\upsilon_F |eE|$; and the probability of the transition is

$$P_{LZ} = \exp\left(-\pi \hbar \upsilon_F k_y^2 / |eE|\right). \quad (3)$$

---

[3] The excitation of graphene optical emission by direct current at much smaller fields was studied in [29,30].

In a general case, within the framework of the Landau-Zener theory the probability of the transition is found by the WKB method as an asymptotic estimate. At the same time, in the work [24] the probability of the transition (3) was found by reducing the Dirac equation to the equation for a parabolic cylinder for which the WKB estimate coincides with the result of the exact solution. We will assess the characteristic time of the transition and the characteristic size of the region in momentum space where the Landau-Zener transitions are significant. It is natural to define the characteristic size of the transition region along the $k_y$ axis by the relation $\delta k_y = \int_{-\infty}^{+\infty} P_{LZ} dk_y = \sqrt{e\hbar^{-1} \upsilon_F^{-1} E}$. The corresponding size of $\delta k_x$ along the $k_x$ axis will be assessed by doubling the minimal width of the energy gap $\Delta W_{min}$ corresponding to $k_x^2 \approx 3 k_y^2$. Making use of $|k_y| \approx \delta k_y / 2$ we obtain $\delta k_x \approx \delta k_y \sqrt{3/2} \approx \sqrt{3eE/2\hbar\upsilon_F}$. With allowance for Eq.(2), the time of quasi-particle flight through the region of electron-hole pair production in phase space may be estimated to be $\delta t_{LZ} \approx \hbar \delta k_x / eE \approx \sqrt{\hbar/eE\upsilon_F}$ [4]. For the $E \approx 100–300$ kV/cm fields, this time is $\delta t_{LZ} \approx 5 – 8$ fs. Thus, for a typical THz pulse duration not less than a hundred fs, the transitions may be regarded to be almost instantaneous.

Changes in the surface density of conduction electrons $N_c$ as a result of interband transitions are defined as

$$\dot{N}_c = \int_{-\infty}^{+\infty} P_{LZ} (\Pi_{x;\upsilon} - \Pi_{x;c}) dk_y, \quad (4)$$

where $\Pi_{x;c,\upsilon} = \eta g |\dot{k}_x| n_{c,\upsilon}$ are the field-initiated particle fluxes in k-space along the $k_x$ axis in the conduction and valence bands, respectively. In the expressions for $\Pi_{x;c,\upsilon}$ fluxes, $n_{c,v}(\mathbf{k})$ are the populations that should be specified at the boundary determined above the region of the Landau-Zener transitions. Two terms in the integrand in Eq.(4) correspond to the electron transition from the valence to the conduction band and back[5]; $\eta = 1/4\pi^2$ is 2D density of state over unit area, $g=4$ is the degeneracy factor for graphene. By substituting Eq. (2) and Eq. (3) into (4) we obtain

$$\dot{N}_c = \pi^{-2}\upsilon_F^{-1/2}\left|\hbar^{-1}eE\right|^{3/2}\left(\langle n_v \rangle - \langle n_c \rangle\right). \qquad (5)$$

---

[4] In the theory of transport processes the quantity $\sqrt{\hbar/v_F eE}$ determines the characteristic threshold time above which the process of Schwinger pair production starts to affect the current-voltage characteristic of a graphene sample [25,28].

In Eq.(5) the populations $\langle n_{c,v} \rangle$ correspond to average values for the quasi-particles pulled by the electric field to the boundary of the Landau-Zener transitions in $k$-space in accordance with the equation of motion (2). For $\langle n_v \rangle = 1$ and $\langle n_c \rangle = 0$ (undoped system), Eq. (5) corresponds to the Schwinger expression for the pair production rate [23,25]. Given $\langle n_v \rangle = \langle n_c \rangle$, from Eq.(5) follows complete compensation of interband exchange processes. At the initial moment of time, this regime corresponds to the quasi-particle energies less than the Fermi energy; it is the reason why the effect of pair production does not influence the current-voltage characteristic of strongly doped graphene [26,27]. At the same time, from Eq. (2) it follows that for the Fermi energy $|W_F| = \hbar v_F k_F$ the region of equal populations is carried away by the field from the neighborhood of the Dirac cone in momentum space during the time $\delta t_F \approx k_F/|\dot{k}_x| = \hbar k_F/eE$. For the field amplitudes mentioned above and a typical magnitude of $|W_F| \approx 0.2$ eV, the time $\delta t_F$ does not exceed several femtoseconds. This time is much shorter than the indicated characteristic duration of the THz pulse; hence, for such intense fields, we can set in Eq. (5) $\langle n_v \rangle = 1$ and $\langle n_c \rangle = 0$, even for relatively heavy initial doping and obtain

$$N_c \approx \pi^{-2}\upsilon_F^{-1/2}\left|\hbar^{-1}eE\right|^{3/2}\Delta t_{\text{eff}}, \qquad (6)$$

where $\Delta t_{\text{eff}}$ is the effective duration of a pulse having amplitude $E$.

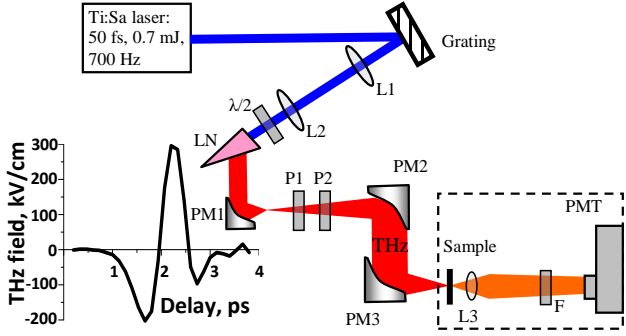

**Fig. 2.** Experimental setup: L1, L2, L3 – lenses; λ/2 – λ/2 plate; LN – LiNbO$_3$ crystal; PM1, PM2, PM3 – parabolic mirrors; P1, P2 – THz polarizers; F – optical filters; PMT – photomultiplier. Inset: time profile of THz pulse.

---
[5] Strictly speaking, the $\Pi_{x;c,v}$ fluxes in Eq.(4) should be multiplied by the Pauli blockade factor $(1 - n_{v,c})$; however, all the $n_c n_v$ products reduce in this case.

Also, for applicability of the relation (6) the characteristic times $\delta t_{LZ}$ and $\delta t_F$ found above should be much less than the typical relaxation times. According to [12,13, 16-22], typical scattering times of Dirac fermions in graphene may be tens of femtoseconds, thermalization may amount from a hundred to several hundreds of femtoseconds, and cooling may be fractions of picoseconds. At the same time, from the results of the work [22,32] also follows feasibility of anisotropic relaxation, when fermion scattering is slower than energy thermalization. However, in any case we are concerned with times significantly greater than $\delta t_{LZ}$ and $\delta t_F$. The recombination time may exceed the thermalization and cooling times several-fold. Anyhow, even the "fastest" recombination occurs at least not faster than thermalization and cooling (see [20]).

Thus, the hierarchy of characteristic relaxation times corresponds to the applicability of the ballistic model of electron-hole breakdown under the action of strong THz pulses. Note that the estimate (6) gives a fairly good description of the results on THz-induced carrier density presented in [12] even neglecting the decisive role of impact ionization.

## 2. Experimental results

We measured the number of optical photons (a wavelength of 340-600 nm) emitted from the graphene sample under the action of a THz pulse. A Li N bO$_3$ crystal irradiated by the Ti-sapphire femtosecond laser (Spitfire, Spectra-Physics) was used as a source of THz radiation. The duration of the optical pulses was 50 fs, energy 1 mJ, central wavelength 795 nm, and repetition rate 700 Hz. The technique of tilted intensity front in a nonlinear Li N bO$_3$ crystal was used to generate THz pulses [33]. The experimental setup is shown in Fig. 2.

The generated THz radiation was collected and transported by means of a telescope consisting of off-axis parabolic mirrors PM1 and PM2 with effective focal lengths (EFL) of 2.5 cm and 19 cm, respectively, and was focused on the sample by the parabolic mirror PM3 (EFL = 5 cm). The diameter of the THz spot on the sample was $\approx 500\,\mu m$ (field amplitude FWHM). The maximum electric field of THz radiation was 300 kV/cm for the THz pulse energy of 0.4 μJ. The characteristic time profile of THz pulses is presented in the inset in Fig. 2. The THz polarizers P1 and P2 were used for THz attenuation.

In our experiments we used a monolayer CVD graphene on a borosilicate glass substrate [34]. As the graphene is deposited on glass, substrate induced inhomogeneity at the graphene-oxide interface gives rise to p-type doping with Fermi energy over 200 meV [12]. Optical emission from the graphene was collected from the solid angle $\Delta o = 0.3$ sr to the photomultiplier (PMT, Hamamatsu R4220P) connected to the photon counting system. A BG39 color filter was placed in front of PMT to eliminate the leakage 800-nm light. A set of color filters placed before PMT was used in the experiment when the spectra of optical emission from graphene were investigated.

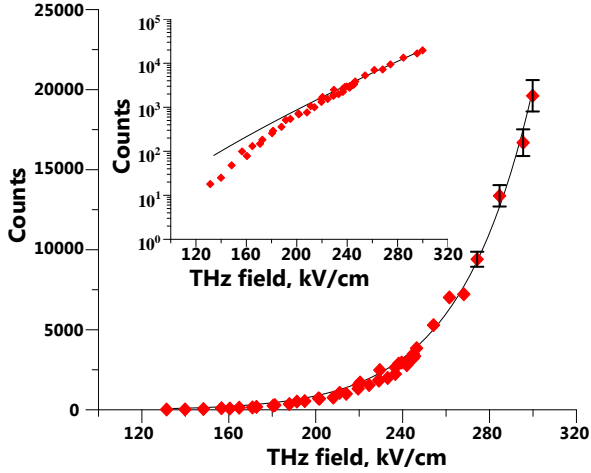

**Fig. 3.** Number of photons emitted by graphene sample as a function of incident terahertz field. The solid line is a fit based on Eqs.(8)-(10) from Sec. 3 ($E_{THz} = 300$ kV/cm corresponds to dimensionless variable $\varepsilon = 1$). Inset: the same dependence on logarithmic scale.

For the THz field values over 100 kV/cm, optical emission from graphene was detected in the 340-600 nm range. The dependence of the number of graphene-emitted photons accumulated during $6 \cdot 10^4$ laser pulses on the terahertz field magnitude is plotted in Fig. 3. The rise of the optical emission by nearly 3 orders of magnitude was observed with the increase of $E_{THz}$ by a factor of 2. The solid curves in Fig. 3 correspond to the ballistic ionization model (see Sec. 3). No optical emission was observed from the glass substrate without graphene.

The spectrum of optical emission from graphene was retrieved using a set of color filters taking into consideration the spectral response of our detection system and the spectral characteristics of the color filters. The emission spectra for the values of the THz field of 300, 250 and 206 kV/cm are shown in Fig. 4 (it is problematic to reliably retrieve the spectrum for fields with smaller amplitudes because of a small number of emitted photons).

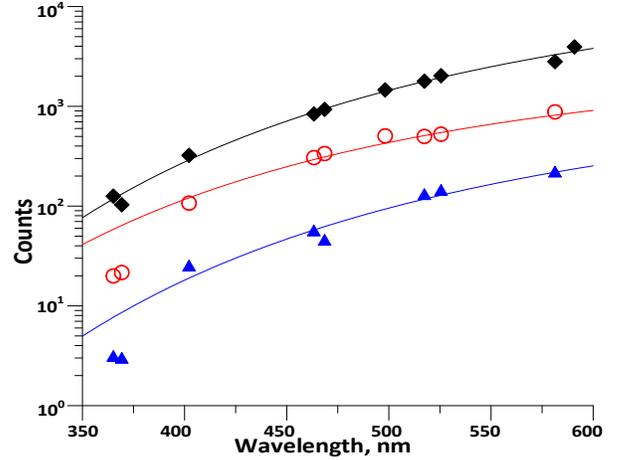

**Fig.4.** Measured spectrum of THz field-excited optical emission. Black rhombs – $E_{THz} = 300$ kV/cm, $T = 0.25$ eV; red circles – $E_{THz} = 250$ kV/cm, $T = 0.28$ eV; blue triangles – $E_{THz} = 206$ kV/cm, $T = 0.25$ eV. The solid curves correspond to thermal spectrum in its Wien's part $\hbar\omega \gg T$.

The solid curves presented in Fig. 4 were plotted assuming equilibrium spectrum: the number of emitted photons in Wien's part of the spectrum $P_0(\omega,T) \propto \omega^2 \exp(-\hbar\omega/T)$. The electron temperature retrieved under the assumption of equilibrium spectrum for the data given in Fig. 4 remains almost unchanged (T=0.25-0.28 eV) to measurement accuracy, while the optical emission increases several tens of times. The effect of saturation of the dependence of temperature on field magnitude may be connected with the sharp decrease of phonon relaxation time with increasing electron energy starting at about 0.2-0.25 eV reported in [35].

**3. Interpretation of experimental data and discussion**

As follows from the data presented in the previous section, with the growth of THz field amplitude from 200 to 300 kV/cm, optical emission increases several tens of times, although the temperature remains unchanged. Such a fast growth of emission intensity despite a nearly constant temperature suggests that not only electron heating but also electron transition to the conduction band under the action of THz pulse are significant in this case. The data about the hierarchy of characteristic times of different processes presented in **Section 1** allows describing the dependence of optical emission intensity on THz pulse amplitude using a simple model of bi-Fermi distribution [26]. Independent chemical potentials for the conduction and valence bands: $n_{c,\upsilon} = \left\{ \exp\left[\left(\pm\hbar\upsilon_F k - \mu_{c,\upsilon}\right)/T\right] + 1 \right\}^{-1}$ correspond to this model for which population inversion occurs in the $\hbar\upsilon_F k < |\mu_\upsilon|$ region. However, as was shown in [20], such inversion at times of order cooling time (fractions of ps) is possible even with a most pessimistic assessment of recombination time.

It is worthy of notice that in the works [22,32] it was pointed to a possibility of anisotropic relaxation, when angular scattering of hot carriers in the conduction band occurs slower than energy thermalization. We will take this into consideration in the optical emission calculation by introducing the anisotropy parameter $\Lambda \leq 1$ for the distribution in the region $2\upsilon_F k \sim \omega$. This parameter is the ratio of the characteristic angular distribution width to $2\pi$ (see **Supplementary Material**, Eq. (S21); $\Lambda = 1$ corresponds to the isotropic case).

Within the framework of the model of bi-Fermi distribution, the spectral photon flux over unit solid angle is given by Eq. (S21) (**Supplementary Material**) that may be represented in the form

$$P_\omega = \Lambda P_0(\omega, T) \cdot \exp\left(-\frac{\mu_c - \mu_\upsilon}{T}\right), \tag{7}$$

where $P_0(\omega, T)$ corresponds to thermal spectrum in its Wien's part. The explicit dependence on the magnitude of THz field in Eq.(7) is contained only in expressions for chemical potentials $\mu_{c,\upsilon}$. The values of $\mu_{c,\upsilon}$ are determined by surface density $N_c$ of the electrons that have passed from the conduction to the valence band under the action of a THz pulse (see **Supplementary Material**, Eqs. (S23)-(S25)). Using for $N_c$ the expression (6) we obtain equations defining chemical potentials

$$-2Li_2\left(-e^{\frac{\mu_c}{T}}\right) = b\varepsilon^{3/2},$$

$$-2Li_2\left(-e^{\frac{\mu_\upsilon}{T}}\right) = b\varepsilon^{3/2} + \frac{\mu_\upsilon^2(0)}{T^2}, \tag{8}$$

where $Li_2(-e^\zeta) = -\int_0^\infty \frac{x\,dx}{\exp(x-\zeta)+1}$ is the second-order polylogarithmic function (Fermi-Dirac integral) [36], $\varepsilon = E_{THz}/E_{THz;max}$ is the relative magnitude of the field, $b = \frac{\sqrt{\hbar}}{\pi T^2}\Delta t_{eff}\left(e\upsilon_F E_{THz;max}\right)^{\frac{3}{2}}$, and $\mu_\upsilon(0) < 0$ is the initial value of chemical potential for a $p$-doped degenerate system. Note that allowance for possible anisotropy of fermion distribution does not affect the qualitative character of the $P_\omega(\varepsilon)$ dependence. We can readily verify that this dependence is retained within the framework of the model with different temperatures $T_{c,\upsilon}$ for different bands.

The total number of detected photons is defined by

$$N_{det}(\varepsilon) = N_0 \cdot \exp\left(\frac{\mu_c - \mu_\upsilon}{T}\right). \tag{9}$$

Under the condition $b\varepsilon^{3/2} \gg 1, \mu_\upsilon^2(0)/T^2$, from Eq.(8) follows a simple asymptotic dependence $(\mu_c - \mu_\upsilon)/T \approx 2\varepsilon^{3/4}\sqrt{2b}$. For $N_0$ we will take the expression that follows from Eq. (S22) (**Supplementary Material**):

$$N_0 = (\pi\alpha)\Upsilon\Lambda\left(\frac{\Delta o}{4\pi}\right)\frac{\Delta t_{em} A T \Omega^2}{\pi^2 c^2 \hbar},$$

$$\Omega^2 = \int X(\omega)\omega^2 \exp\left(-\frac{\hbar\omega}{T}\right)\frac{\hbar d\omega}{T}. \tag{10}$$

Here, $\pi\alpha = \pi e^2/c\hbar \approx 0.023$ is a standard coefficient of interband absorption in graphene, $X(\omega)$ is the spectral efficiency of our detection system[6], $\Upsilon = 60 \cdot 10^3$ is the total number of THz shots, and $A$ is the effective area of the emitting single layer. The expression (10) corresponds to the sum of photons of both polarizations emitted into a relatively small solid angle $\Delta o$ in the direction normal to the single layer of graphene during time $\Delta t_{em}$ (this time should, evidently, be chosen to be of order cooling time of quasi-particles). Note that for $N_{det}(r) \propto \exp\left\{[\varepsilon(r)]^{3/4}\sqrt{2b}\right\}$ with characteristic spatial scale of field localization $L_{THz}$, the size of the emission region is $L_{em} \approx L_{THz}/\sqrt{2b}$. For the parameter $b=16$ (corresponding to the experimental dependence) and $L_{THz} \approx 500\,\mu m$ we obtain $L_{em} \approx 80\,\mu m$ and $A \approx 5 \cdot 10^{-5}\,cm^2$.

The expressions (8)-(10) describe the relative dependence of emission on THz field amplitude quite well. Only in the region of sufficiently weak fields $E_{THz}/E_{THz;max} < 0.5$ the difference of the theoretical curve from the experimental data evidently indicates appearance of the dependence of temperature $T$ on the field (in this region the emission is attenuated by about 2 orders of magnitude as compared to the emission at $E_{THz} \approx E_{THz;max}$). Agreement for absolute values (see Fig. 4) is attained for the parameter values $b=16$, $\mu_\upsilon(0)/T \approx 1$ and $N_0 = 0.4$. For the solid angle $\Delta o = 0.3\,sr$, $E_{THz;max} = 300\,kV/cm$, $\Omega^2 = 3 \cdot 10^{24}\,sec^{-2}$; these parameters correspond, e.g., to the following reasonable magnitudes: $\Delta t_{eff} \approx \Delta t_{em} \approx 400\,fs$, $T = 0.2\,eV$, $\Lambda = 0.2$ and spectral efficiency of PMT ~1% at 600 nm wavelength.

A more detailed comparison of theory and experiment demands kinetic calculations taking into account a complex form of the THz pulse[7] (see the inset in Fig. 2) and the difference of fermion distribution from the simple model used above. We intend to carry out such calculations in the near future. Nevertheless, we believe that the conclusion about the determining impact of the Landau-Zener transitions on the process of optical emission excitation is justified already on the basis of the measurements and their interpretation presented in this work.

---

[6] The function $X(\omega)$ also takes into account a) the effect of emission attenuation of the dipole placed on the surface of a substrate with dielectric permeability $\varepsilon_d$ ($\propto \varepsilon_d^{-1}$ factor); b) the beam refraction effect, which also gives the $\propto \varepsilon_d^{-1}$ factor at detection of photons that have passed through a dielectric.

## ACKNOWLEDGMENTS

The authors are grateful to A.A. Belyanin and I.D. Tokman for fruitful discussions and to N.B. Krivatkina for help in preparing the manuscript.

---

[7] These calculations may provide a more accurate definition of $\Delta t_{\text{eff}}$ and elucidate its dependence on relative field amplitude $\varepsilon$.

**Supplementary Material. Spontaneous emission of graphene monolayer**

**1. Basic equations**

Consider massless Dirac fermions in the emission field using the Hamiltonian of the system

$$\hat{H} = \sum_{v,q} \hbar\omega_q \left( \hat{c}^\dagger_{vq} \hat{c}_{vq} + \frac{1}{2} \right) + \sum_{sk} W_s(k) \hat{a}^\dagger_{sk} \hat{a}_{sk} + \sum_{ss'kk'} \hat{V}_{ss'kk'} \hat{a}^\dagger_{sk} \hat{a}_{s'k'}. \quad (S1)$$

Here, $\hat{c}^\dagger_{vq}$ and $\hat{c}_{vq}$ are the operators of creation and annihilation of Fock photon states $|n_{vq}\rangle$ corresponding to wave vector $q$, the subscript $v$ stands for photon polarization, and $\omega_q^2 = c^2 q^2$. The fermions are described by the creation and annihilation operators $\hat{a}^\dagger_{sk}$ and $\hat{a}_{sk}$ corresponding to massless Dirac states $|k, s\rangle$ [1]:

$$|k,s\rangle = \frac{e^{ikr}}{\sqrt{2A}} \begin{pmatrix} s \\ e^{i\theta(k)} \end{pmatrix}, \quad W_s(k) = s\hbar v_F k, \quad (S2)$$

where $A$ is the area of a monolayer lying in the $xy$-plane, $k$ is the 2D wave vector of a quasi-particle, the indices $s = \pm 1$ correspond to the eigenfunctions for the conduction and valence bands, respectively, and $\theta(k)$ is the angle between the quasi-momentum and the $x$-axis. Summation over $k$ in (S1) formally implies summation over spin states and valleys.

In (S2) $\hat{V}_{ss'kk'}$ is the matrix element of the interaction operator [1], which in the case of a quantum field must depend on $\hat{c}^\dagger_{vq}$ and $\hat{c}_{vq}$ operators:

$$\hat{V} = -\frac{1}{c} \hat{j}\hat{A}, \quad (S3)$$

where $\hat{j} = -e v_F \hat{\sigma}$ is the operator of current, $\hat{\sigma} = x_0 \hat{\sigma}_x + y_0 \hat{\sigma}_y$, $\hat{\sigma}_x$ and $\hat{\sigma}_y$ are the Pauli matrices, $\hat{A}$ is the vector potential operator

$$\hat{A} = \sum_{v,q} \sqrt{\frac{2\pi c^2 \hbar}{V \omega_q}} \cdot \left( e_v \hat{c}_{vq} e^{-i\omega_q t + iqz} + e_v^* \hat{c}^\dagger_{vq} e^{i\omega_q t - iqz} \right), \quad (S4)$$

$V$ is quantization volume. Let $v = S, P$ correspond to standard $S$ and $P$ polarizations of photons, i.e., the unit vector $e_S$ lies in the monolayer plane, and the unit vector $e_P$ lies in the plane formed by vector $q$ and the normal to the monolayer. The direction of vector $q$ is specified by the angle $\Theta_q$ relative to the normal to the monolayer and by the angle $\Phi_q$ relative to the $x$-axis in the $xy$-plane. For such polarization vectors, the case $\Theta_q = 0$ is degenerate: we can consider for this case photons polarized along the $x$- and $y$-axes, i.e., we can take $v = x, y$.

Expressions for the matrix elements $\hat{V}_{ss'kk'}$ are obtained taking into consideration the reasonable condition $q \ll k$. To this approximation we will have

$$\hat{V}_{+1+1kk'} \approx e v_F \sum_q \sqrt{\frac{2\pi \hbar}{V \omega_q}} \times \left[ \sin(\Phi_q - \theta) \cdot \left( \delta_{k(k'+q_\perp)} \hat{c}_{Sq} e^{-i\omega_q t} + \delta_{k(k'-q_\perp)} \hat{c}^\dagger_{Sq} e^{i\omega_q t} \right) + \right.$$
$$\left. + \cos\Theta_q \cdot \cos(\Phi_q - \theta) \cdot \left( \delta_{k(k'+q_\perp)} \hat{c}_{Pq} e^{-i\omega_q t} + \delta_{k(k'-q_\perp)} \hat{c}^\dagger_{Pq} e^{i\omega_q t} \right) \right] \quad (S5)$$

$$\hat{V}_{+1-1kk'} \approx ie\upsilon_F \sum_q \sqrt{\frac{2\pi\hbar}{V\omega_q}} \times \left[\cos(\Phi_q-\theta)\cdot\left(\delta_{k(k'+q_\perp)}\hat{c}_{Sq}e^{-i\omega_q t}+\delta_{k(k'-q_\perp)}\hat{c}^\dagger_{Sq}e^{i\omega_q t}\right)-\right.$$
$$\left.-\cos\Theta_q\cdot\sin(\Phi_q-\theta)\cdot\left(\delta_{k(k'+q_\perp)}\hat{c}_{Pq}e^{-i\omega_q t}+\delta_{k(k'-q_\perp)}\hat{c}^\dagger_{Pq}e^{i\omega_q t}\right)\right]$$
(S6)

$$\hat{V}_{-1-1kk'}=-\hat{V}_{+1+1kk'},\quad \hat{V}_{-1+1kk'}=-\hat{V}_{+1-1kk'},\tag{S7}$$

where $q_\perp$ is the vector component $q$ in the plane of graphene monolayer: $q_\perp=q|\sin\Theta_q|$.

## 2. Probability of interband spontaneous transition

Consider a spontaneous transition between the states $|k',+1\rangle \to |k,-1\rangle$ accompanied by photon emission with $v$-type polarization. For the probability of such a transition per unit time $w_v$ we can right away use the gold Fermi rule [2]:

$$w_v=\frac{2\pi}{\hbar}\int d\Pi_f \left|V_{v,fi}\right|^2 \delta(W_i-W_f-\hbar\omega)\tag{S8}$$

The integration $\int d\Pi_f$ in (S8) is done over all finite states of the system, $i$ is the initial state, and $\hbar\omega$ is the photon energy. In this case we have $W_i=\hbar\upsilon_F k'$, $W_f=-\hbar\upsilon_F k$; the matrix element $V_{v,fi}$ is $V_{v,fi}=\langle 1_{vq}|\hat{V}_{-1+1kk'}|0_{vq}\rangle$, where $|n_{vq}\rangle$ is the corresponding Fock state, and we obtain

$$\left|V_{S,fi}\right|^2=e^2\upsilon_F^2\cos^2\left[\Phi_q-\theta(k)\right]\frac{2\pi\hbar}{V\omega_q}\delta_{k(k'-q_\perp)},\tag{S9a}$$

$$\left|V_{P,fi}\right|^2=e^2\upsilon_F^2\cos^2\Theta_q\sin^2\left[\Phi_q-\theta(k)\right]\frac{2\pi\hbar}{V\omega_q}\delta_{k(k'-q_\perp)}.\tag{S9b}$$

The emitted photon frequency is determined by the condition $W_i-W_f=\hbar\omega$. Taking into consideration the relations $\omega=2\upsilon_F(|k|+|k+q_\perp|)$ and $q=c/\omega$ and the inequality $q\ll k$, we obtain

$$\omega\approx\frac{2\upsilon_F k}{1+\frac{\upsilon_F}{c}\sin\Theta_q\cos[\Phi_q-\theta(k)]}.\tag{S10}$$

Further, with the expression for the density of states of photons with given polarization in (S8) taken into account, we have $d\Pi_f=(2\pi c)^{-3}V\omega^2 d\omega d\Omega$, where $d\Omega=\sin\Theta_q d\Theta_q d\Phi_q$ is the element of the solid angle in the direction of the photon wave vector $q$. The substitution of (S9a,b) into (S8) yields the expression for the probability of spontaneous emission into a unit solid angle:

$$w_{\Omega;S}=\frac{e^2\upsilon_F^2\omega}{2\pi\hbar c^3}\cos^2\left[\Phi_q-\theta(k)\right],\tag{S11a}$$

$$w_{\Omega;P}=\frac{e^2\upsilon_F^2\omega}{2\pi\hbar c^3}\cos^2\Theta_q\sin^2\left[\Phi_q-\theta(k)\right],\tag{S11b)s}$$

where the emitted photon frequency is specified by the relation (S10).

### 3. Summation over electron states

Let us sum the expressions (S11a,b) over 2D electron states:

$$\sum_k (...) \Rightarrow \frac{gA}{4\pi^2}\int_\infty (...)d^2k = \frac{gA}{4\pi^2}\int_0^{2\pi}d\theta\int_0^\infty (...)kdk, \qquad (S12)$$

where $g=4$ is the factor of degeneracy with respect to spin states and valleys. Making use of the approximate relation $\omega = 2\upsilon_F k$ following from (S10), via (S11a,b) and (S12) we obtain an expression for spectral photon fluxes $P_{\omega\Omega;S,P}$:

$$\begin{bmatrix} P_{\omega\Omega;S} \\ P_{\omega\Omega;P} \end{bmatrix} = \frac{A\omega^2}{2^3\pi^4 c^2}(\pi\alpha)\int_0^{2\pi}\begin{bmatrix} \cos^2(\Phi_q-\theta) \\ \cos^2\theta_q \sin^2(\Phi_q-\theta) \end{bmatrix}\{n_c(k,\theta)[1-n_\upsilon(k,\theta)]\}_{k=\omega/2\upsilon_F} d\theta, \qquad (S13)$$

where $\pi\alpha = \pi e^2/c\hbar \approx 0.023$ is a standard coefficient of interband absorption in graphene, $n_{c,\upsilon}(k,\theta) = \langle \hat{a}^\dagger_{\pm 1k}\hat{a}_{\pm 1k}\rangle$ are average occupation numbers of photon states (populations), $1-n_\upsilon$ is the Pauli blockade factor. For fermion distributions isotropic with respect to angle $\theta$, from (S13) follows

$$\begin{bmatrix} P_{\omega\Omega;S} \\ P_{\omega\Omega;P} \end{bmatrix} = \frac{A\omega^2}{(2\pi)^3 c^2}\begin{bmatrix} 1 \\ \cos^2\theta_q \end{bmatrix}\pi\alpha\{n_c(k)[1-n_\upsilon(k)]\}_{k=\omega/2\upsilon_F}. \qquad (S14)$$

### 4. Emission of equilibrium ensemble of quasi-particles

Consider the Fermi distribution

$$n_{c,\upsilon} = \left[\exp\left(\frac{\pm\hbar k\upsilon_F + \mu}{T}\right)+1\right]^{-1}. \qquad (S15)$$

The substitution of (S15) into (S14) yields

$$\begin{bmatrix} P_{\omega\Omega;S}(T) \\ P_{\omega\Omega;P}(T) \end{bmatrix} = \begin{bmatrix} 1 \\ \cos^2\Theta_q \end{bmatrix}\frac{A\omega^2}{(2\pi)^3 c^2}\cdot\frac{\pi\alpha}{\left[\exp\left(\frac{\hbar\omega/2+\mu}{T}\right)+1\right]\cdot\left[\exp\left(\frac{\hbar\omega/2-\mu}{T}\right)+1\right]}. \qquad (S16)$$

To confirm correctness of calculating luminescence we will show that (S16) corresponds to the Kirchhoff law. Indeed, following [1] we can readily obtain an expression for monolayer optical thickness determined by interband absorption for isotropic distributions $n_{c,\upsilon}(k)$ at arbitrary value of $\Theta_q$:

$$\begin{bmatrix} \Gamma_S \\ \Gamma_P \end{bmatrix} = \begin{bmatrix} \cos^{-1}\theta_q \\ \cos\theta_q \end{bmatrix}\pi\alpha[n_\upsilon(k)-n_c(k)]_{k=\omega/2\upsilon_F} \qquad (S17)$$

(for $\theta_q \ll 1$ and $n_c = 0$, $n_\upsilon = 1$ we obtain $\Gamma_{S,P} = \pi\alpha$). For the equilibrium distribution (S15), from (S17) follows

$$\begin{bmatrix} \Gamma_S(\omega,T) \\ \Gamma_P(\omega,T) \end{bmatrix} = \begin{bmatrix} \cos^{-1}\Theta_q \\ \cos\Theta_q \end{bmatrix}\pi\alpha\frac{\left[\exp\left(\frac{\hbar\omega}{T}\right)-1\right]}{\left[\exp\left(\frac{\hbar\omega/2+\mu}{T}\right)+1\right]\cdot\left[\exp\left(\frac{\hbar\omega/2-\mu}{T}\right)+1\right]}. \qquad (S18)$$

Comparison of (S16) and (S18) gives

$$P_{\omega\Omega;S,P}(T) = \Gamma_{S,P}(\omega,T)A\cos\Theta_q P_0(\omega,T), \qquad (S19)$$

where $P_0(\omega,T) = \dfrac{\omega^2}{(2\pi)^3 c^2} \cdot \left[\exp\left(\dfrac{\hbar\omega}{T}\right)-1\right]^{-1}$ is the spectral flux of photons with fixed polarization through unit area in unit solid angle [2]. The expression (S19) explicitly expresses the Kirchhoff law.

### 5. Spontaneous emission of bi-Fermi distribution

Let there occur in each band a distribution with its own chemical potential

$$n_{c,v} = \left[\exp\left(\dfrac{\pm\hbar k\upsilon_F - \mu_{c,v}}{T}\right)+1\right]^{-1}. \tag{S20}$$

We will find the spectral photon flux into the lens placed above the emitting area at an angle $\Theta_q = 0$ and collecting photons of both polarizations from a relatively small solid angle $\Delta o$. Consider Wien's region of the spectrum $(\hbar\omega/2 \mp \mu_{c,v} \gg T)$ and use the expressions (S13) and (S20). We will qualitatively allow for possible anisotropy of high-energy ($2\upsilon_F k \sim \omega$) carrier distribution assuming $n_c \neq 0$ in the sector $-\Delta\theta/2 < \theta < \Delta\theta/2$ to obtain

$$P_\omega \approx \left(\dfrac{\Delta o}{4\pi}\right)\dfrac{\Lambda A \omega^2}{\pi^2 c^2}\pi\alpha \cdot \exp\left(-\dfrac{\hbar\omega}{T}\right)\exp\left(\dfrac{\mu_c - \mu_v}{T}\right), \tag{S21}$$

where $\Lambda \leq 1$ is the anisotropy parameter. Within the framework of the considered simple model $\Lambda = \Delta\theta/2\pi$.

Instead of the above simplest model of anisotropic angular distribution, a standard method of Legendre polynomial expansion may be used. Taking into account the first two terms of the Legendre series, under the condition $n_c \leq 1$ we obtain the expression

$$n_c = \left[\exp\left(\dfrac{\hbar k\upsilon_F - \mu_c}{T}\right)+1\right]^{-1} \times \left[1-\lambda(1-\cos\theta)\right]. \tag{S22}$$

It can be readily ascertained that the use of (S22) also leads to (S21), where $1-\lambda = \Lambda$.

### 5. Expressions for chemical potentials $\mu_{c,v}$

Let us find the relationship between the chemical potential $\mu_c$ and the surface density $N_c$ of the electrons that have passed to the conduction band from the valence band under the action of THz pulse. For the degeneracy factor $g=4$ and density of states $\eta = 1/4\pi^2$ we obtain

$$\int_0^\infty \dfrac{2x\,dx}{\exp\left(x-\dfrac{\mu_c}{T}\right)+1} = \pi\left(\dfrac{\hbar\upsilon_F}{T}\right)^2 N_c. \tag{S23}$$

The chemical potential in the valence band is determined by surface density of vacancies $N_h$:

$$\int_0^\infty \dfrac{2x\,dx}{\exp\left(x+\dfrac{\mu_v}{T}\right)+1} = \pi\left(\dfrac{\hbar\upsilon_F}{T}\right)^2 N_h, \tag{S24}$$

where

$$N_h \approx \dfrac{\mu_v^2(0)}{\pi\hbar^2 \upsilon_F^2} + N_c, \tag{S25}$$

$\mu_\upsilon(0) < 0$ is the initial value of chemical potential for the *p*-doped degenerate system. Eqs. (S23), (S24) are valid when most of fermions belong to the isotropic part of distribution function.